\documentstyle[epsfig,11pt]{article}
\def\be{\begin{eqnarray}}
\def\ee{\end{eqnarray}}

\def\half{{\textstyle \frac{1}{2}}}

\def\roughly#1{\mathrel{\raise.3ex\hbox{$#1$\kern-.75em%
\lower1ex\hbox{$\sim$}}}}

\begin{document}

\renewcommand{\thefootnote}{\arabic{footnote}}
\setcounter{footnote}{0}

\vskip 0.4cm
\hfill {\bf KIAS-P99047}\\
$\mbox{}$\hfill {\bf FZJ-IKP(TH)-1999-27}\\

\hfill {\today}
\vskip 1cm

\begin{center}
{\LARGE\bf Dense QCD : \\[3mm]
Overhauser or BCS Pairing ?}

\date{\today}

\vskip 1cm
Byung-Yoon Park$^{a,b}$\footnote{E-mail:
bypark@chaosphys.chungnam.ac.kr}, Mannque
Rho$^{a,c}$\footnote{E-mail: rho@spht.saclay.cea.fr},
Andreas
Wirzba$^{d,e}$\footnote{E-mail: a.wirzba@fz-juelich.de}
and Ismail  Zahed$^{a,d}$\footnote{E-mail:
zahed@nuclear.physics.sunysb.edu} 

\end{center}

\vskip 0.5cm

\begin{center}

$^a$
{\it School of Physics, Korea Institute for Advanced Study,
Seoul 130-012, Korea}

$^b$
{\it Department of Physics, Chungnam National University,
Taejon 305-764, Korea}

$^c$
{\it Service de Physique Th\'eorique, CE Saclay,
91191 Gif-sur-Yvette, France}


$^d$
{\it Department of Physics and Astronomy,
SUNY-Stony-Brook, NY 11794, U.\,S.\,A.}

$^e$
{\it FZ J{\"u}lich, Institut f\"ur Kernphysik (Theorie),  D-52425
  J{\"u}lich, Germany}

\end{center}

\vskip 0.5cm

\begin{abstract}
We discuss the Overhauser effect (particle-hole pairing)
versus the BCS effect (particle-particle or hole-hole pairing)
in QCD at large quark density. In weak coupling and to leading
logarithm accuracy, the pairing energies can be estimated exactly.
For a small number of colors, the BCS effect
overtakes the Overhauser effect, while for a large number of
colors the opposite takes place, in agreement with a recent
renormalization group argument. In strong coupling with
large pairing energies, the Overhauser effect may be dominant for
any number of colors, suggesting that QCD may crystallize into
an insulator at a few times nuclear matter density, a situation
reminiscent of dense Skyrmions. The Overhauser effect is dominant
in QCD in 1+1 dimensions, although susceptible to quantum effects. 
It is sensitive to temperature in all dimensions.

\end{abstract}
\newpage

\renewcommand{\thefootnote}{\#\arabic{footnote}}
\setcounter{footnote}{0}


\centerline{\bf 1. Introduction}
\vskip 1cm

Quantum chromodynamics (QCD) at high density, relevant to the
physics of the early universe, compact stars and relativistic
heavy ion collisions, is presently attracting a renewed attention
from both nuclear and particle theorists. Following an early
suggestion by Bailin and Love~\cite{LOVE}, it was recently stressed
that at large quark density, diquarks could condense into a color
superconductor~\cite{ALL}, with potentially interesting
and novel phenomena such as color-flavor locking,
chiral symmetry breaking, parity violation, color-flavor anomalies,
and superqualitons.

At large density, quarks at the edge of the Fermi surface
interact weakly thanks to asymptotic freedom. However,
the high degeneracy of the Fermi surface causes perturbation
theory to fail. As a result, particles can pair and condense
at the edge of the Fermi surface leading to energy gaps.
Particle-particle and hole-hole pairing (BCS effect) have been
extensively studied recently~\cite{LOVE,ALL}. Particle-hole
pairing  at the opposite edges of the Fermi surface
(Overhauser effect)~\cite{OVERHAUSER} has received little attention
with the exception of an early variational study by Deryagin,
Grigoriev and Rubakov for a large number of colors~\cite{DGR92},
and a recent renormalization group argument in~\cite{SON2}.
The scattering amplitude between a pair of particles at the opposite
edges of the Fermi surface peaks in the forward direction, a situation
reminiscent of the forward enhancement in Compton and Bhabha scattering.

In retrospect, it is surprising that the Overhauser effect in QCD has
attracted so little attention. In fact, the Schwinger model~\cite{SCHWINGER}
shows that when a uniform external charge density is applied,
the electrons respond by screening the external charge and inducing
a charge density wave, a situation analogous to a Wigner crystal
~\cite{SUSSKIND, SCHAPOSNIK,KAO}.
Similar considerations apply to
QCD in 1+1 dimensions \cite{SCHAPOSNIK}.
In 3+1 dimensions, dense Skyrmion calculations with
realistic chiral parameters yield a 3-dimensional Wigner-type crystal
with half-Skyrmion symmetry at few
times nuclear matter density~\cite{SKYRMION,MANTON}. At these densities, Fermi
motion is expected to be overtaken by the classical interaction~\cite{COHEN}.
A close inspection of these results shows the occurrence of scalar-isoscalar,
pseudoscalar-isovector and vector-isoscalar charge density waves in an
ensemble of dense Skyrmions.

In this paper we will show that in dense QCD,
the equations that drive the particle-hole instability at
the opposite edge of a Fermi surface resemble
those that drive the
particle-particle or hole-hole instability in the scalar-isoscalar channel,
modulo phase-space factors. In section 2 we motivate and
derive a Wilsonian action around the Fermi surface. In section 3 we
obtain expressions for the energy densities and pertinent gaps in
the $0^+$ channel with screening, thereby generalizing the original
results in~\cite{DGR92}. In section 4, we analyze the decoupled equations
for large chemical potential without screening. The effects of screening
for arbitrary $N_c$ as well as temperature are discussed in section 5, in
overall agreement with a recent renormalization group
argument~\cite{SON2}.
In section 6, we discuss the Overhauser effect in QCD in lower dimensions.
Our conclusions and suggestions are given in section 7.

\vskip 1.5cm
\centerline{\bf 2. Effective Action at the Fermi Surface}
\vskip 1cm

To compare the Overhauser effect to the BCS effect, we will
construct a Wilsonian effective action by integrating out the quark modes
around the Fermi surface, in the presence of smooth bilocal fields.
An alternative would be the quantum action~\cite{NRZ}. At large chemical
potential, most of the Fermi surface is Pauli-blocked, so the quasiparticle
content of the theory is well described by such an action. Incidentally,
our analysis should provide a useful alternative to a brute-force lattice
QCD analysis. Indeed, an effective formulation of lattice QCD along the lines
of the heavy-quark formalism is possible and will be discussed
elsewhere~\cite{PREPARATION}.

The starting point in our analysis
is the appropriate QCD action in Euclidean space with
massless quarks
\begin{equation}
   S = \int d^4x\, \left[
       \frac14 (F_{\mu\nu}^a)^2
        + \bar{\psi} (\gamma_\mu \partial_\mu - \gamma_4 \mu) \psi
       - i J^a_\mu  A^a_\mu \right]\;,
 \label{S1}
\end{equation}
and the colored current
\begin{equation}
   J_\mu^a = g \bar{\psi} \gamma_\mu \frac{\lambda^a}{2} \psi\;.
   \label{S2}
\end{equation}
In Euclidean space, our conventions are such that
the $\gamma$-matrices are hermitean with
$\{\gamma_\mu,\gamma_\nu\}=2\delta_{\mu\nu}$.
%
For sufficiently
large $\mu$, we will assume $g^2 N_c \ll 1 $. We have omitted
gauge-fixing terms and  ghost-fields. In what follows, we will
analyze (\ref{S1}) in the one-loop approximation with the gluon field
in the Feynman gauge. The approximation, as we shall show below, is
equivalent to the resummation of the ladder graphs in the
particle-particle or particle-hole graphs. The effects of screening
will be dealt with by minimally modifying the gluon propagator, ignoring
for simplicity vertex corrections as in~\cite{ALL}. The issue of gauge
fixing dependence will be briefly discussed at the end.

In the one-loop approximation with screened gluons, the induced action is
\begin{equation}
    S_\psi = \frac{g^2}{2} \int d^4x\, d^4y\,
     J^a_\mu (x) {\cal D}_{\mu\nu} (x-y) J^a_\nu(y)
     + \int d^4x \,\bar{\psi}\, \tilde{\partial}_\mu \gamma_\mu \psi\;,
 \label{S3}
\end{equation}
where $\tilde{\partial}_\mu =
(\partial_1,\partial_2,\partial_3, \partial_4 - \mu)$.
The screened gluon propagator $\{{\cal D}_{\mu\nu}\}
=({\cal D}_E, {\cal D}_M)$ is
\begin{equation}
    {\cal D}_{E,M}(x-y) =  \int \frac{d^4 q}{(2\pi)^4}\,
                   \frac{1}{q^2 + m_{E,M}^2} e^{-iq\cdot (x-y)} \; .
 \label{S4}
\end{equation}
Perturbative arguments
give $m_E^2/(g\mu)^2=m_D^2/(g\mu)^2\approx
N_f/2\pi^2$ and $m_M^2/m_D^2\approx \pi{|q_4|/|4{\bf q}|}$, where
$m_D$ is the Debye mass, $m_M$ is the magnetic screening generated
by Landau damping and $N_f$ the number of flavors~\cite{LeBELLAC}~\footnote{Throughout
we will refer to $m_M$ abusively as
the magnetic screening mass.}. Nonperturbative arguments suggest
$m_E^2, m_M^2\rightarrow m_*^4/q^2$ \cite{ZWAN} where
for simplicity, the difference between electric
and magnetic channels is ignored.
We expect $\Lambda_{QCD}\ll m_*< m_E$ in the case $N_c=3$, as lattice simulations
for the gluon propagator at finite $\mu$ are not yet available.
We note that the perturbative screening vanishes at large $N_c$.

To proceed further with (\ref{S3}) we need to Fierz rearrange
the $JJ$ term in (\ref{S3}).
This is equivalent to summing ladder graphs with relevant
quantum numbers. Specifically,
\begin{eqnarray}
  J^a_\mu(x) {\cal D}_{\mu\nu}(x-y) J^a_\nu(y)
  &=&g^2\sum_{\cal O} \,\, {\cal C}_{\cal O}
    \left[\bar{\psi}(x) {\bf M}_{\cal O} \psi (y)\right]
            \, {\cal D}(x-y)\,
    \left[\bar{\psi}(y) {\bf M}_{\cal O}\psi(x)\right] \nonumber \\
  &+& g^2\sum_{{\cal O'}} \,\,{\cal C}_{{\cal O}'}
    \left[\bar{\psi}(x) {\bf M}_{{\cal O}'} \psi^c (y)\right]
            \, {\cal D}(x-y)\,
    \left[\bar{\psi}^c(y) {\bf M}_{{\cal O}'}\psi(x)\right]
  \label{Fierzing}
\end{eqnarray}
with ${\cal C}_{O}=-1/9$ and ${\cal C}_{C}=+1/36$ for
the operators
\begin{eqnarray}
  \left[ \bar{\psi}(x)\, {\bf M}_O\, \psi(y) \right] & = &
   \bar{\psi}_{\alpha,a,i}(x)\, \delta_{\alpha\beta}\,\delta_{ab}\delta_{ij}
  \,\psi_{\beta,b,j}(y)\;, \nonumber\\
  \left[ \bar{\psi}(x)\, {\bf M}_{C}\, {\psi}^{c}(y)\right]  & = &
  \bar{\psi}_{\alpha,a,i}(x)\, (\gamma_5)_{\alpha\beta}\,
  \varepsilon^I_{ab}\varepsilon^I_{ij}\, C\bar{\psi}^{T}_{\beta,b,j}(y)\;,
 \label{S7}
\end{eqnarray}
respectively, with $N_f=N_c=3$.
These quantities involve matrices active in color $(a,b,\cdots)$, flavor $(i,j,\cdots)$
and Dirac space $(\alpha,\beta,\cdots)$. ${\bf M}_O$ is the vertex generator for
particle-hole pairing in the $0^+$ channel
(i.e., Overhauser), while ${\bf M}_C$ is the vertex generator
for particle-particle and hole-hole pairing in the color-flavor locked
(CFL) channel (i.e., BCS). Only these two operators will be retained below,
unless specified otherwise. The gluon-propagator in matter is
\begin{equation}
  {\cal D}(x-y) =
   \textstyle \frac12{\cal D}_E(x-y)+\frac12{\cal D}_M(x-y) \;.
  \label{S6}
\end{equation}
The weightings follow from minimal substitution in matter
with 2 electric and 2 magnetic modes. We note that the
present Fierzing is particular, since it selects
solely the ${\bf 1}_c$ in the $\overline{q}q$ channel and the
$\overline{\bf 3}_c$ in the $qq$ channel~\cite{CAHILL}.
For arbitrary $N_c\geq 3$ and $N_f\geq 2$, the coefficients
$-\textstyle \frac{1}{9}$ and $\textstyle \frac{1}{36}$ become, respectively,
$-\frac{1}{2}(1-\frac{1}{N_c})\cdot\frac{1}{N_f}$ and
$\frac{1}{2 N_c}\cdot\frac{1}{2}\cdot\frac{1}{ \min(N_c,N_f)}$, where
the single factors refer, in turn, to the results of
the color Fierzing, the flavor Fierzing and, of course only for the second
expression, the Fierzing
related to color-flavor
locking~\footnote{At least partially, even if $N_f \neq N_c$, a locking
can be achieved by Fierzing the
antisymmetric tensor
in color
times the corresponding one in flavor
into the tensor ${\bf M}_{ai,bj}
=\delta_{ai}\delta_{bj}-\delta_{aj}\delta_{bi}$
with the pertinent weight ${1}/{\min(N_c,N_f)}$
in the combined color-flavor space.
The latter operator  has $\half n(n-1)$
eigenvalues $+1$, $\half n(n+1)\mbox{$-$}1 $ eigenvalues $-1$, one eigenvalue
$n-1$, and $N_c\times N_f- n^2$ eigenvalues 0, where $n\equiv
\min(N_c,N_f)$. Thus, in the BCS case, the fermion determinant
(\ref{S15}) acquires the color-flavor weight
$2n(n-1)$, while   the corresponding  value in the Overhauser case is
the standard $N_c N_f$ factor. \label{footlocking}}.
To compare to the more conventional decompositions through
${\bf 3}_c\times \overline{\bf 3}_c = {\bf 1}_c + {\bf 8}_c$
for $\overline{q} q$ and
${\bf 3}_c\times {\bf 3}_c=\overline{\bf 3}_c +{\bf 6}_c$ for $qq$,
with respective weights
$-\frac{1}{2}(1-\frac{1}{N_c^2})\cdot\frac{1}{N_f}$ and
$\frac{N_c+1}{4 N_c}\cdot\frac{1}{2}\cdot \frac{1}{\min(N_c,N_f)}$,
we introduce also the vertex generator~\cite{ALL}
\begin{equation}
   \left[ \bar{\psi}(x)\, {\bf M}_{B}\, {\psi}^{c}(y)\right]   =
   \bar{\psi}_{\alpha,a,i}(x)\, (\gamma_5)_{\alpha\beta}\,
   (\lambda_2)_{ab} (\tau_2)_{ij} C\bar{\psi}^{T}_{\beta,b,j}(y)
  \label{S8} \; .
\end{equation}
We note that (\ref{S8}) does not lock color and flavor as it
stands;
a color-flavor locking  as described in
footnote~\ref{footlocking} still has to be
performed, such that finally the corresponding coefficient becomes
${\cal C}_B=(N_c+1)/\{8N_c \min(N_f,N_c)\}$~\footnote{In fact,
expression (\ref{S8}) is the operator considered in
Ref.~\cite{PISARSKI,SchaferWilczek} where
color and flavor are uncoupled and only the two-flavor case is
considered (see also \cite{SON2,SON1}). Thus
the corresponding coefficient is just  $(N_c+1)/4N_c$, since the flavor-Fierzing
factor can be ignored, as it eventually cancels against a
corresponding factor resulting from the
fermion determinant. Note, furthermore, that our
color-flavor coupling
scheme is
different from the one recently introduced in Ref.~\cite{SchaferNew}
for arbitrary numbers of flavors.}.
This brings about the important issue of whether Fierzing
is a unique operation on 4-fermi interactions. The answer is
no~\cite{EBERT,LANGFELD}. This nonuniqueness would
of course not be important if an all-order calculation were to be performed
for any Fierzing set, but is of course relevant for truncated calculations
as is the case in general. Each Fierzing corresponds to
summing a specific class of ladder diagrams in the energy density,
see e.g.~\cite{EBERT,LANGFELD}.

Introducing a hermitian bilocal field $\Sigma(x,y)$
and a non-hermitian bilocal field
$\Gamma(x,y)$, we may linearize the Fierzed form
of the $JJ$ term by using the Hubbard-Stratonovich transformation, e.g.
\begin{eqnarray}
  \lefteqn{\exp \left(\textstyle\frac{g^2}{18} \displaystyle
   \int d^4x\, d^4y\, \left[\bar{\psi}(x)\psi(y)\right]
  {\cal D}(x-y)
   \left[\bar{\psi}(y)\psi(x)\right] \right) }\nonumber\\
   &=&\int d\Sigma(x,y)\,
   \exp \left(-S_\Sigma
   - \int d^4x\, d^4y \,\bar{\psi}(x)\, \Sigma(x,y)\, \psi(y) \right)
 \label{S9}
\end{eqnarray}
with
\begin{equation}
   {S}_\Sigma =
   \frac{9}{2g^2} \int d^4x \,d^4y\,
   \frac{\left| \Sigma(x,y)\right|^2}{{\cal D}(x-y)}
 \label{S10}
\end{equation}
and similarly for $\Gamma$.
As a result, the action in the quark fields is linear
and the functional integration can be performed. The result is
the following effective action for the bilocal fields
\begin{equation}
  {S} = {S}_\Sigma
       +  {S}_\Gamma -\half \mbox{Tr}{\ } \mbox{ln} {\bf F} \; ,
 \label{S12}
\end{equation}
where
%
\begin{equation}
  {\bf F}= \left(\begin{array}{cc} \left\{\gamma \cdot {\partial}-\mu\gamma_4
\right\}\delta(x-y) + {\bf M}_O \Sigma(x,y)  &
   i\Gamma^\dagger(x,y) C^T {\bf M}_{C} \\ iC^T\Gamma(x,y) {\bf M}_{C}  &
\left\{\gamma \cdot {\partial}+ \mu\gamma_4\right\}
\delta(x-y) + {\bf M}_O \Sigma(x,y)
 \end{array}\right)  .
 \end{equation}
The factor of $1/2$ in (\ref{S12}) is due to the occurrence of $\psi$
and $\psi^c$ through the Fierzing into ${\bf 1}_c$ and ${\bf
3}_c$~\cite{CAHILL}. This renders naturally the
Gorkov formalism applicable to the present problem even at $\mu=0$.
Note that ${\bf M}_O = {\bf 1}_C \times {\bf 1}_F \times {\bf 1}_D$ and
${\bf M}_C = \varepsilon_C^I \times \varepsilon_F^I \times \gamma_5$,
with the subscripts $C,F,D$ short for color, flavor and Dirac.
We should stress that
the effective action (\ref{S12}) is general. The third term is the
Hartree contribution of the quarks to the ground state energy at
large chemical potential, while the first two terms remove the
double counting in the potential (i.e., Fock terms).

To analyze the Overhauser
and BCS effects in parallel, we make simplifying ans\"atze for the
bilocal auxiliary fields. Since the unscreened gluon interaction in
both cases peak in the forward direction, we may choose
\begin{eqnarray}
  \Sigma(x,y) &\!=\!& 2
   \cos\left[ P_\mu \left(\frac{x_\mu + y_\mu}{2} \right) \right]
  \sigma(x-y) =2 \cos\left[ P_\mu \left(\frac{x_\mu + y_\mu}{2} \right) \right]
  \int \frac{d^4 q}{(2\pi)^4}\, e^{-i q\cdot(x-y)} F(q)\; ,
  \nonumber  \\
  \Gamma(x,y) &\!=\!& 2 \cos\left[
  P_\mu \left(\frac{x_\mu - y_\mu}{2} \right) \right]
  g(x-y) =2 \cos\left[ P_\mu \left(\frac{x_\mu - y_\mu}{2} \right) \right]
  \int \frac{d^4 q}{(2\pi)^4}\, e^{-i q\cdot(x-y)} G(q)\; ,\nonumber\\
 &&
\label{S14}
\end{eqnarray}
where ${P_\mu} = ({\bf P}_F,0)$ and $|{\bf P}_F|=2\mu$. ${\bf P}_F$
points in the original direction of one of the quark.
$F(q)$ and $G(q)$ are even functions,
$F(q)$ is real, since $\Sigma (x,y)=\Sigma (y,x)^\ast$,
and $G(q)$ is complex, since $\Gamma^\dagger(x,y)=\Gamma(y,x)^\ast$.
The relative momentum $q$ satisfies $|q|\leq |P/2|=\mu$.
The bilocal field $\Gamma$ characterizes a BCS pair of zero total momentum.
$\Sigma$ characterizes a wave of total momentum $2\mu$. This is
the optimal choice for the momentum of the standing wave for which the holes
contribute coherently to the wave formation. As a result the gap opens up at
the Fermi surface, with $\mu$ as the divide between particles and
holes. In both cases, the pairing
involves a particle and/or hole at the opposite sides of the Fermi surface.
Indeed, in terms of (\ref{S14}) the linear terms in the bilocal fields are
\begin{eqnarray}
 \lefteqn{\int d^4x\, d^4y\, \bar\psi(x)\Sigma(x,y) \psi(y)} \nonumber\\
  &=&V_4 \int \frac{d^4q}{(2\pi)^4}\,
\left[\bar\psi\left(-\frac{P}{2}\mbox{+}q\right) \,F(q)\,
  \psi\left(\frac{P}{2}\mbox{+}q\right) +
  \bar\psi\left(\frac{P}{2}\mbox{+}q\right) \,F(q)\,
      \psi\left(-\frac{P}{2}\mbox{+}q\right)\right]
\end{eqnarray}
(see Ref.\cite{DGR92}) and
\begin{eqnarray}
  \lefteqn{
\half\int d^4x\, d^4y\,
  \left[
  \bar\psi^{c}(x) i\gamma_5\Gamma(x,y) \psi(y)
  +\bar\psi(x)\Gamma^\dagger(x,y)i\gamma_5 \psi^{c}(y)\right] }
 \nonumber \\
\! &\!\!=\!\!&\! \half V_4\!\int\! \frac{d^4q}{(2\pi)^4}\left[
  \psi^{T}\left(-\frac{P}{2}\mbox{$-$}q\right)
  C i\gamma_5 G(q)\,\psi\left(\frac{P}{2}\mbox{+}q\right)
+  \bar\psi\left(\frac{P}{2}+q\right)
  iG^\ast(q)\gamma_5 C  \psi^{T}\left(-\frac{P}{2}\mbox{$-$}q\right)
  \right. \nonumber \\
&& \;\mbox +\left.
\psi^{T}\left(\frac{P}{2}\mbox{$-$}q\right)\,
  C i\gamma_5G(q)\,\psi\left(-\frac{P}{2}\mbox{+}q\right)
+  \bar\psi\left(-\frac{P}{2}\mbox{+}q\right)\,
  iG^\ast(q)\gamma_5 C  \psi^{T}\left(\frac{P}{2}\mbox{$-$}q\right)
  \right],
\end{eqnarray}
where $V_4$ is the 4-volume.

Following~\cite{DGR92}, we introduce fermion fields
$\psi(\pm P/2 +q)$ and
$\psi^c (\pm P/2 -q)$~\footnote{Note that we define
$\psi^{c}(k)\equiv C\bar\psi^T(k)$ in terms of the
Euclidean charge conjugation operator $C=\gamma_4\gamma_2$, whereas in
Ref.\cite{Pisarski:1999cn} $\psi^{c}(k)\equiv C\bar\psi^T(-k)$.
As usual,  $T$ stands for ``transposed''.}
that are
independent integration variables in the relevant region of the
momentum $|q|\ll |{\bf P}|/2$. Hence, the quark contribution
around the Fermi surface can be integrated. The result is~\cite{USLATER}
\begin{eqnarray}
 \det{\bf F}={\rm exp}\left(V_4\,{\rm Tr\,\,ln}
  \left|{\begin{array}{cccc}
 -i\widetilde Q_{+,\mu}{\bf \sigma}_\mu &
   F(q) &
   iG^\ast(q)\,\,{\bf M}_{C}&
    {\bf 0}
\\
  F(q)   &
 -i\widetilde Q_{-,\mu} {\bf\bar \sigma}_\mu  &
    {\bf 0}  &
  -iG^\ast(q)\,\,{\bf M}_{C}
\\
  -iG(q)\,\,{\bf M}_{C}&
     {\bf 0} &
  -i\widetilde Q_{+,\mu}^{*}{\bf \bar\sigma}_\mu   &
   F(-q)
 \\
   {\bf 0}  &
  iG(q)\,\, {\bf M}_{C} &
  F(-q)\,\, &
  -i\widetilde Q_{-,\mu}^{*} {\bf \sigma}_\mu
 \end{array}} \right|\,\right) \; ,
 \label{S15}
\end{eqnarray}
where $Q_{\pm}\equiv \pm\frac{P}{2}+q$
and $ \widetilde Q_{\pm}  \equiv\left ({\bf Q}_{\pm}\,\,,\,
Q^4_{\pm} -i\mu\, \right)$. For each entry in momentum space $q$,
the determinant in (\ref{S15})
is over an $(8\cdot N_c\cdot N_f)\times (8\cdot N_c\cdot N_f)$-matrix.
The matrices  $\sigma_\mu=(i\,{\vec \sigma}\,,\, {\bf 1})$ and
$\bar\sigma_\mu=(-i\,{\vec \sigma}\,,\, {\bf 1})$ are
defined in terms of the usual
Pauli matrices ${\vec \sigma}$.
The detailed analysis of the coupled
problem (\ref{S15}) with the full Fermion determinant will
be discussed elsewhere~\cite{USLATER}.

\vskip 1.5cm
\centerline{\bf 3. Gap Equations}
\vskip 1cm

A qualitative understanding of the Overhauser effect versus the BCS
effect can be achieved by studying the phases separately, and then
comparing their energy densities at large quark density.
Setting $G=0$ yields, for the Overhauser pairing, an energy density
\begin{equation}
  \frac{{S}_{\Sigma}}{9V_4} =
  \frac{1}{g^2} \int d^4x\, \frac{|\sigma(x)|^2}{{\cal D}(x)}
  - 2\int\frac{d^4 q}{(2\pi)^4}\, \ln \left[\frac{
  \tilde{Q}_+^2 \tilde{Q}_-^2 + 2 F^2\, \tilde{Q}_+\tilde{Q}_- + F^4 }
  {\tilde{Q}_+^2 \tilde{Q}_-^2}\right]
  \equiv {\cal S}_{\mbox{\scriptsize pot},\Sigma}
 + {\cal S}_{\mbox{\scriptsize kin},\Sigma}
 \label{S16}
\end{equation}
which is in agreement with the result derived originally in~\cite{DGR92}.
Setting $F=0$ yields, for the BCS pairing, an energy density
\begin{equation}
  \frac{{S}_\Gamma}{36V_4} =
  \frac{1}{g^2} \int d^4x\, \frac{|g (x)|^2}{{\cal D}(x)}
  - \frac 23\int\frac{d^4 q}{(2\pi)^4}\, \ln \left[\frac{
  \tilde{Q}_+^{2} \tilde{Q}_+^{\ast 2}
  + 2 |G|^2\, \tilde{Q}_+ \tilde{Q}_+^\ast + |G|^4 }
   {\tilde{Q}_+^2 \tilde{Q}_+^{*2}}\right]
   \equiv {\cal S}_{\mbox{\scriptsize pot},\Gamma}
                    + {\cal S}_{\mbox{\scriptsize kin},\Gamma}
 \label{S17}
\end{equation}
which is similar to (\ref{S16}). In writing the last equation
we have assumed that $|G(q\pm P)| \approx |G(\pm\mu)|\approx 0$.
The gap equations for both cases follow by variation. The result is
\begin{equation}
  F(p) = 2g^2 \int \frac{d^4 q}{(2\pi)^4}\, {\cal D}(p-q)
  \left( \frac{2F(q)\left(\tilde{Q}_+ \tilde{Q}_- + F^2(q)\right)}
  {\tilde{Q}_+^2 \tilde{Q}_-^2 + 2F^2(q)\, \tilde{Q}_+ \tilde{Q}_- + F^4(q)}
   \right)
 \label{S18}
\end{equation}
for the Overhauser gap, and
\begin{equation}
  G(p) = {\textstyle \frac{2}{3}}g^2
  \int \frac{d^4 q}{(2\pi)^4}\, {\cal D}(p-q)
  \left( \frac{2G(q)\left(\tilde{Q}_+ \tilde{Q}_+^\ast + |G(q)|^2\right)}
  {\tilde{Q}_+^2 \tilde{Q}_+^{*2} +
  2|G(q)|^2\, \tilde{Q}_+ \tilde{Q}_+^\ast + |G(q)|^4}\right)
 \label{S19}
\end{equation}
for the BCS gap.
If we were to use the antisymmetric vertex operator
(\ref{S8}) then we would have $2g^2/3\rightarrow 4g^2/3$. For the latter, we
have checked that the results
(\ref{S18}-\ref{S19}) agree with the Bethe-Salpeter derivation in
the ladder approximation to order $\mu^0$.
In our notations, the leading order effects are of
order $\mu$, the next to leading order
effects are of order $\mu^0$
and the next-to-next to leading order effects are
of order $\mu^{-1}$.
For the screened gluon propagator we have the alternatives
\begin{eqnarray}
 && {\cal D}(q) = \textstyle\frac12 \displaystyle \frac{1}{q^2 + m_E^2}
  + \textstyle\frac12 \displaystyle \frac{1}{q^2 + m_M^2} \; ,
\nonumber\\
 &&{\cal D}(q) = \textstyle\frac12 \displaystyle \frac{1}{q^2 + im_*^2}
  + \textstyle\frac12 \displaystyle \frac{1}{q^2 -im_*^2}\; ,
\end{eqnarray}
for the perturbative and nonperturbative assignments respectively.

The present construction is valid for an arbitrary number of colors
with or without screening, thereby generalizing the original analysis
in~\cite{DGR92}. The outcome can be analyzed variationally, numerically
or even analytically to leading logarithm accuracy.
Using the following momentum decomposition around the fixed Fermi
momentum $P$ at the Fermi surface,
\begin{equation}
  q_{||} = \frac{{\bf P}\cdot{\bf q}}{|{\bf P}|}\,, \hskip 3em
  {\bf q}_{\perp} = {\bf q} - q_{||}\frac{{\bf P}}{|{\bf P}|}\,,
\end{equation}
and assuming that the relevant values of the
amplitudes of the bilocal fields are small
(i.e., $F, |G| \ll \mu$), we may further simplify
the kinetic part in the energy densities
eqs.(\ref{S16}-\ref{S17}). Specifically,
\begin{eqnarray}
&&{\cal S}_{\mbox{\scriptsize kin},\Sigma} \approx
  - 2\int\frac{d^4 q}{(2\pi)^4}\, \ln \left[\frac
{q_{||}^2 + F^2(q) +
\left\{q_4 + \frac {q^2}{2i\mu}\right\}^2}
{q_{||}^2 +
\left\{q_4 + \frac {q^2}{2i\mu}\right\}^2}
\right] \; ,
\nonumber\\
&& {\cal S}_{\mbox{\scriptsize kin},\Gamma}\approx
  - \frac 23\int\frac{d^4 q}{(2\pi)^4}\, \ln \left[\frac
{q_{4}^2 + |G(q)|^2 +
\left\{q_{||} + \frac {q^2}{2\mu}\right\}^2}
{q_{4}^2 +
\left\{q_{||} + \frac {q^2}{2\mu}\right\}^2}
\right] \; .
 \label{S17x}
\end{eqnarray}
The simplified gap equations are
\begin{equation}
  F(p) \approx 2g^2 \int \frac{d^4q}{(2\pi)^4}\, {\cal D}(p-q)
  \left[ \frac{F(q)}{q_{||}^2 + F^2(q) +
(q_4 + \frac {q^2}{2i\mu})^2}\right]
 \label{S22}
\end{equation}
and
\begin{equation}
  G(p) \approx \frac 23 g^2 \int \frac{d^4q}{(2\pi)^4}\, {\cal D}(p-q)
  \left[ \frac{G(q)}{q_4^2 + |G(q)|^2 +
(q_{||} +\frac {q^2}{2\mu})^2}\right].
 \label{S23}
\end{equation}
For both pairings, the simplified energy densities
$\overline{S}_{\Sigma,\Gamma}$ at their respective extrema
are
\begin{eqnarray}
  \frac{\overline{S}_{\Sigma}}{9V_4} &\approx& 2
  \int \frac{d^4 q}{(2\pi )^4}\, \left(\frac 12 F\partial_F-1\right)
  \,\ln\left(1+ \frac{F^2(q)}{q_{||}^2 + (q_4+\frac {q^2}{2i\mu})^2}\right)
 \; ,
 \nonumber\\
  \frac{\overline{S}_{\Gamma}}{36V_4} &\approx& \frac 23
  \int \frac{d^4 q}{(2\pi)^4} \, \left(\frac 12 |G|\partial_{|G|}-1\right)
  \,\ln\left(1+ \frac{|G(q)|^2}{q_4^2 +(q_{||} +\frac {q^2}{2\mu})^2}\right)
  \; . \nonumber
\end{eqnarray}
We now proceed to evaluate $F,G$ to leading logarithm accuracy.

\vskip 1.5cm
\centerline{\bf 4. Unscreened Case: Large $N_c$}
\vskip 1cm

In this section we consider the gap equations (\ref{S22}-\ref{S23})
in the absence of screening. In the perturbative regime, we note
that $m_{E,M}\sim 1/N_c$, and this approximation may be somehow
justified in large $N_c$~\cite{DGR92}. Hence,
\begin{equation}
  F(p) \approx 2g^2 \int \frac{d^4q}{(2\pi)^4}\, \frac 1{(p-q)^2}
  \left[ \frac{F(q)}{q_{||}^2 + F^2(q) +
(q_4 + \frac {q^2}{2i\mu})^2}\right]
 \label{S222}
\end{equation}
and
\begin{equation}
  G(p) \approx \frac 23 g^2 \int \frac{d^4q}{(2\pi)^4}\, \frac 1{(p-q)^2}
  \left[ \frac{G(q)}{q_4^2 + |G(q)|^2 +
(q_{||} +\frac {q^2}{2\mu})^2}\right].
 \label{S233}
\end{equation}

For the Overhauser pairing, if we assume the propagator to be static,
the $q_4$ integration can be performed by a contour-integration with
the constraint that
\begin{equation}
|q_{\perp}|^2 \leq 2\mu\epsilon_q \equiv 2\mu \sqrt{q_{||}^2+F^2(q_{||})}\; .
\label{PERP}
\end{equation}
Hence
\begin{equation}
F(p_{||})\approx h^2\int_0^\infty dq_{||}\,\frac{F(q_{||})}{\epsilon_q}\,
{\rm ln}\left( 1+ \frac{2\mu\epsilon_q}{(p_{||}-q_{||})^2}\right)
\label{OVERno}
\end{equation}
with $h^2 =\frac{g^2}{4\pi^2}$. In general, we have
\begin{eqnarray}
h^2 &\equiv& \frac{g^2 N_c}{8\pi^2}\left(1-\frac{1}{N_c}\right)\;,\nonumber\\
h^2 &\equiv& \frac{g^2 N_c}{8\pi^2}\left(1-\frac{1}{N^2_c}\right)\; ,
\label{SH}
\end{eqnarray}
for Fierzing with ${\bf M}_C$ and ${\bf M}_B$~\cite{DGR92} respectively.
Eq.~(\ref{OVERno}) is essentially a one-dimensional `fish-diagram' with
logarithmically running couplings. This feature
is preserved by screening as we will show below,
in agreement with the
recent renormalization group analysis in~\cite{SON2}.
Following \cite{SON1}, the resulting
equations are readily solved  by defining the logarithmic scales
$x\equiv\ln\,(2\mu/p_{||})$, $y\equiv\ln\,(2\mu/q_{||})$,
$x_0\equiv\ln\,(2\mu/F_0)$, and rewriting
\begin{equation}
  F(x)\approx  h^2 \left(2x \int_{x}^{x_0} dy\, F(y)
     -\int_x^{x_0} dy\, y F(y)
     + \int_0^x dy\, y F(y) \right)\; .
 \label{Fintegro}
\end{equation}
Since $F''(x)= -2h^2 F(x)$ with $F(x_0)= -F(0)$, then
$ F(x)=- F_0 \cos(\sqrt{2}h x)$~\cite{DGR92,SON1}. The coefficient $F_0$
follows from $F'(x_0)=0$, with
$\sqrt{2}h x_0=\pi$. Hence $F_0= F(x_0)$ and
$x_0\equiv\ln(2\mu/F_0)= \pi/(\sqrt{2}h)$.
Thus
\begin{equation}
   F_0 \sim 2\mu \exp \left\{-\frac{\pi}{ \sqrt{2}h} \right\}
\label{SGAPE}
\end{equation}
which is exactly the result established in~\cite{DGR92} using
the ${\bf M}_C$ Fierzing and elaborate variational arguments.
Note that the pairing energy $F_0\ll \Lambda_{\perp}\ll \mu$
follows from an exponentially small region in transverse
momentum (\ref{PERP}) as required by momentum conservation,
see Fig.~1b. Typically $\Lambda_{\perp}=\sqrt{2\mu F_0}$ as
originally suggested in~\cite{DGR92}.

For the BCS pairing, the transverse momentum is not restricted
as shown in Fig.~1a. This is best illustrated by noting that the
BCS equation in (\ref{S233}) can be further simplified through the
following substitution
\begin{equation}
  q_{||}+\frac {{\bf q}^2}{2\mu}
  \rightarrow \left|{\bf q}+\frac {\bf P}2\right|-\mu
\; .
\label{APPROX}
\end{equation}
This  amounts to taking into account the effects of curvature
around the fixed Fermi momentum $P/2$ defined by the standing
wave. The trade (\ref{APPROX}) allows for a larger covering of the
Fermi surface, although for $\Lambda_{\perp}=2\mu$ the terms that
are dropped are only subleading for $q_{||}^2\ll q_{\perp}^2$.
We have checked that this substitution does not not affect
our analysis in the leading logarithm approximation.
Shifting momenta to $Q=q+P/2$ and $K=p+P/2$ yields
\begin{equation}
  G(K-P/2) \approx \frac 23 g^2 \int \frac{d^4Q}{(2\pi)^4}\, {\cal D}(K-Q)
  \left[ \frac{G(Q-P/2)}{Q_4^2 + |G(Q-P/2)|^2 +
(|{\bf Q}|-\mu)^2}\right]\; .
 \label{S24x}
\end{equation}
For a constant gap, the $Q$-integration diverges logarithmically.
As most of the physics follows from $|{\bf Q}|=\mu$, this divergence
can be regulated~\cite{PISARSKI}, with no effect on the leading-logarithm
estimate of the pairing energy. Hence,
\begin{equation}
G(p_{||})\approx  h_*^2\int_0^\infty dq_{||}\,\frac{G(q_{||})}{\epsilon_q}\,
{\rm ln}\left( 1+ \frac{4\mu^2}{(p_{||}-q_{||})^2}\right)
\label{BCSno}
\end{equation}
with $\epsilon_q=\sqrt{q_{||}^2+|G(q_{||})|^2}$ following from the
contour integration over $Q_4$. The prefactor reads $h_*^2=g^2/12\pi^2$, and in general
~\footnote{In~\cite{SON1} and footnote 1 of \cite{SON2}
color and flavor are uncoupled.
Hence $h_*^2=\frac{g^2}{8\pi^2}(1+\frac 1{N_c})$. In fact, this value
can also be reproduced by  the Fierzing with
${\bf M}_B$ for the special case $N_f=3$.}
\begin{eqnarray}
h_\ast^2 &=& \frac {g^2}{8\pi^2} \ \left( \frac{2}{N_c}\right)
   \frac{\min\, (N_f,N_c)-1}{2}\;,\nonumber\\
h_\ast^2 &=& \frac {g^2}{8\pi^2} \
\left(1+\frac{1}{N_c}\right) \frac{\min\,(N_f,N_c)-1}{2} \; ,
\label{SH*}
\end{eqnarray}
corresponding to Fierzing with
${\bf M}_C$ and ${\bf M}_B$ respectively. Notice the
similarity between (\ref{OVERno}) and (\ref{BCSno}),
especially in the one-dimensional reduction of the
equations.
In terms of the logarithmic scales, the BCS equation
reads~\cite{SON1}
\begin{equation}
  G(x)\approx  2h_*^2 \left(x \int_{x}^{x_0} dy\, G(y)
       + \int_0^x dy\, y G(y) \right)\; .
 \label{Gintegro}
\end{equation}
Since $G''(x)= -2h_*^2 G(x)$ with $G(0)=0$, then
$ G(x)= G_0\sin(\sqrt{2}h_* x)$. The coefficient $G_0$
follows from $G'(x_0)=0$  with
$\sqrt{2}h_* x_0=\pi/2$. Hence $G_0=G(x_0)$ and, because of
$x_0\equiv\ln(2\mu/G_0)$,
\begin{equation}
   G_0 \sim 2\mu \exp \left\{-\frac{\pi}{2\sqrt{2}h_*} \right\} \; .
\label{SGAPEBCS}
\end{equation}
Note that $G_0$ is enhanced relative to  $F_0$, if  $N_c=3$.
They both become comparable for $N_c\geq 4$ in the ${\bf M_C}$-Fierzing case
with the Overhauser
effect dominating at large $N_c$~\footnote{In the ${\bf M}_B$-Fierzing
case, the Overhauser effect only dominates in the large $N_c$
limit, if $N_f<\half N_c$.\label{f5}},
as originally suggested in~\cite{DGR92}.

We note that the  $i$ in (\ref{S22})
(Overhauser) versus no $i$ in (\ref{S23}) (BCS) stems from the kinematical
difference between the two pairings, hence a difference in the phase-space
integration  due to momentum conservation as shown in Fig.~1a and 1b.
In weak coupling,
both gaps are exponentially small. The energy budget can be
assessed by noting that the phase space volumes are of order:
$\mu^2\,G_0$ (BCS) and $\mu F_0^2$ (Overhauser). Hence, the
energy densities are
\begin{equation}
\frac{{S}_{\Sigma}}{V_4}\approx -\mu\,\,F_0^3\; ,\qquad\qquad
\frac{{S}_{\Gamma}}{V_4}\approx -\mu^2\,\,G_0^2\; .
\label{budget}
\end{equation}
In weak coupling, the BCS phase is energetically favored up to
$N_c\sim 10$ in the unscreened case and for one standing
wave for the ${\bf M}_C$ Fierzing. Under
${\bf M}_B$ Fierzing, we have an additional constraint on the
number
of flavors, e.g.,
$N_f < {\textstyle\frac {2}{9}} N_c$ for large $N_c$.
Remember that further nestings of the Fermi
surface by particle-hole pairing are still possible as shown in Fig.~1c,
causing a further reduction in ${S}_{\Sigma}/V_4$. A total nesting
of the Fermi surface will bring about $4\pi\mu^2/\Lambda_{\perp}^2\approx
\mu/F_0$ patches, hence ${S}_{\Sigma}/V_4\approx -\mu^2F_0^2$.  The
BCS phase becomes comparable to the Overhauser phase for $N_c\sim 4$
(see, however, footnote \ref{f5}).
Finally, we note that in strong coupling, both gaps are a fraction of $\mu$.

\begin{figure}
\epsfig{file=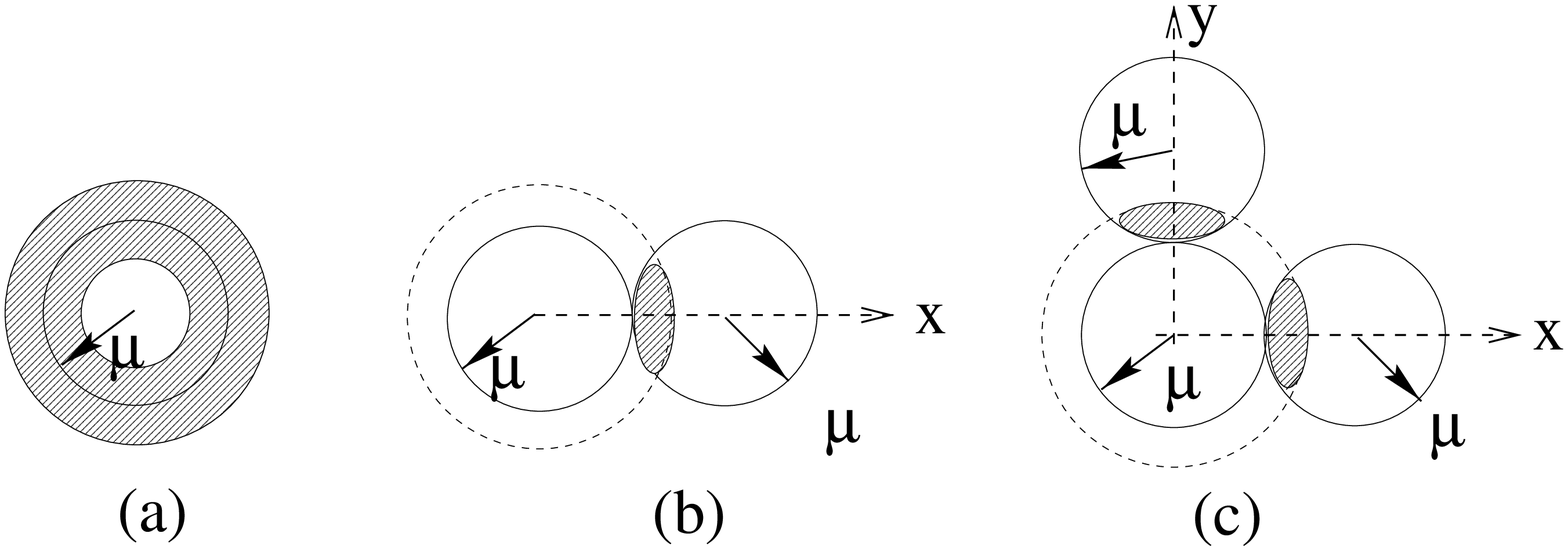,height=2in}
\caption{(a) fraction of the Fermi surface used in BCS pairing;
(b) fraction of the Fermi surface used in the Overhauser pairing with
one standing wave; (c) fractions of the Fermi surface used in the
Overhauser pairing with two orthogonal standing waves.}
\end{figure}

\vskip 1.5cm
\centerline{\bf 5. Screened Case: Finite $N_c$}
\vskip 1cm

In the presence of electric and magnetic screening, which are important
in matter, the situation
changes significantly. While the original variational arguments in
\cite{DGR92} were tailored for the unscreened case, our formulation
which reproduces exactly their unscreened results in the leading
logarithm approximation, generalizes naturally to the screened
perturbative and nonperturbative cases in a minimal way. Indeed, using
(\ref{S22}-\ref{S23}) and the pertinent transverse cutoffs, we obtain for
perturbative screening,
\begin{eqnarray} 
  F(p_{||}) &\!\!\approx\!\!& \frac{h^2}{6}\int_0^{\infty}\! dq_{||}\,
  \frac{F(q_{||})}{\sqrt{q_{||}^2+F^2(q_{||})}}\nonumber\\
   &&\qquad\mbox{}\times
   \,\ln\left\{
   \left(1+\frac{\Lambda_{\perp}^2}{(p_{||}\mbox{$-$}q_{||})^2+m_E^2}\right)^3
   \left(1+\frac{\Lambda_{\perp}^3}
  {|p_{||}\mbox{$-$}q_{||}|^3
 +\frac{\pi}{4}m_D^2|p_{||}\mbox{$-$}q_{||}|}\right)^2\,\right\}  ,
   \nonumber\\
  G(p_{||}) &\!\!\approx\!\!& \frac{h^2_\ast}{6}\int_0^{\infty}\! dq_{||}\,
   \frac{G(q_{||})}{\sqrt{q_{||}^2+|G(q_{||})|^2}}\nonumber\\
  &&\qquad\mbox{}\times
   \,\ln\left\{
   \left(1+\frac{\Lambda_{\perp}^2}{(p_{||}\mbox{$-$}q_{||})^2+m_E^2}\right)^3
   \left(1+\frac{\Lambda_{\perp}^3}{|p_{||}\mbox{$-$}q_{||}|^3
 +\frac{\pi}{4}m_D^2|p_{||}\mbox{$-$}q_{||}|}
    \right)^2\,\right\}  ,\nonumber \\
 & &
 \label{S27}
\end{eqnarray}
and for nonperturbative screening
\begin{eqnarray}
  F(p_{||}) &\approx&  h^2\int_0^{\infty} dq_{||}\,
  \frac{F(q_{||})}{\sqrt{q_{||}^2+F^2(q_{||})}}
   \,\ln\left| 1+\frac{\Lambda_{\perp}^2}{(p_{||}-q_{||})^2+im_*^2}\right|
\; ,
   \nonumber\\
  G(p_{||}) &\approx&  h_\ast^2\int_0^{\infty} dq_{||}\,
   \frac{G(q_{||})}{\sqrt{q_{||}^2+|G(q_{||})|^2}}
   \,\ln\left| 1+\frac{\Lambda_{\perp}^2}{(p_{||}-q_{||})^2+im_*^2}\right|
\nonumber\\
 \label{S28}
\end{eqnarray}
where the transverse cutoffs are $\Lambda_{\perp}=\sqrt{2\mu\epsilon_q}$
(Overhauser) and $\Lambda_{\perp}=2\mu$ (BCS) respectively. The cutoffs
are exactly fixed in weak coupling, and reflect on the fractions
of the Fermi surface used in the pairing.

In the BCS case, the transverse cutoff is large.
Hence $\Lambda_{\perp}>m_E,m_M$ and the
logarithm in (\ref{S27}) may
not be expanded. Dropping 1, we obtain
to leading logarithm accuracy,
\begin{eqnarray}
G_0 \approx \left(\frac{4 \Lambda_{\perp}^6}{\pi m_E^5}\right)\,
  e^{-\frac{\sqrt{3}\pi}{2h_\ast}} \; .
\label{Sgapp}
\end{eqnarray}
The results for the BCS gap is the same as the one reached
in~\cite{SON1,SchaferWilczek,PISARSKI}~\footnote{Modulo
the dimensionless constant $b_0'$ in~\cite{PISARSKI}.}
if we were to Fierz with ${\bf M}_B$ instead of ${\bf M}_C$.
Note that (\ref{Sgapp}) is smaller than (\ref{SGAPEBCS}) as expected.
For nonperturbative screening, the result is
\begin{equation}
G_0\approx \Lambda_{\perp}
e^{-\frac{2}{h_\ast^2}\left\{\ln (1+\Lambda_{\perp}^4/m_*^4)\right\}^{-1}}
\end{equation}
with $\Lambda_{\perp}/m_*=2 \mu/m_*\gg 1$.

In the Overhauser case, the transverse cutoff is reduced
in comparison to the BCS case due to momentum conservation
for fixed 3-momentum for the standing wave. The equation
can be rearranged into the form
\begin{eqnarray}
F(p_{||})&\approx & \frac {h^2}6\int_0^{\infty} dq_{||}\,
\frac{F(q_{||})}{\epsilon_q}\,
{\rm ln}\left( \frac{2\mu\epsilon_q}{(p_{||}-q_{||})^2}\right)\nonumber\\
&&\mbox{} +\frac {5h^2}6\int_0^{\infty} dq_{||}\,
\frac{F(q_{||})}{\epsilon_q}\,
{\rm ln}\left( \frac{2\mu\epsilon_q}{(p_{||}-q_{||})^2+m_E^2}\right)\,,
\label{OVERyes}
\end{eqnarray}
where we have approximated $\pi m_D^2/4\sim m_E^2$ and used static, but
perturbatively screened propagators. The effects of Landau damping through
the magnetic gluons result into an unscreened interaction but with a reduced
strength $h^2\rightarrow h^2/6$. Eq.~(\ref{OVERyes}) can be solved to
leading logarithm accuracy using the logarithmic scales as defined
above. Specifically, for $x<x_m\equiv\ln(2\mu/m_E)$, we get
\begin{equation}
  F(x)\approx  h^2 \left(2x \int_{x}^{x_L} dy\, F(y)
     -\int_x^{x_L} dy\, y F(y)
     + \int_0^x dy\, y F(y) \right)
 \label{Fintegr1}
\end{equation}
as in the unscreened case with $x_L=x_0$, and for $x>x_m$
\begin{equation}
  F(x)\approx  \frac{h^2}6 \left(2x \int_{x}^{x_R} dy\, F(y)
     -\int_x^{x_R} dy\, y F(y)
     + \int_0^x dy\, y F(y) \right) +{\cal C}\; .
 \label{Fintegr2}
\end{equation}
Here $x_{L,R}\equiv \ln(2\mu/F_{L,R})$ and
the constant ${\cal C}$ is given by
\begin{equation}
{\cal C}=\frac {5h^2}{6}\int_0^{\infty}dq_{||}\,
\frac{F(q_{||})}{\epsilon_q}\,
{\rm ln}\left(\frac {2\mu\epsilon_q }{{\rm max}\,(q_{||}^2,m_E^2)}\right)\; .
\label{constant}
\end{equation}
The solution to (\ref{Fintegr1}-\ref{Fintegr2}) is
\begin{equation}
\begin{array}{lclclcl}
 F(x)&=&F_L \cos(\sqrt{2} \,h\,(x-x_L))& \qquad\mbox{for}\quad     & x &
  < & x_m\; ,\nonumber\\
 F(x)&=&F_R \cos(h\,(x-x_R)/\sqrt{3})& \qquad\mbox{for}\quad & x & >
 & x_m \; .
\label{SOLUTION}
\end{array}
\label{SOL}
\end{equation}
We note that for $x<x_m$ or $p_{||}>m_E$, screening can be ignored
to leading logarithm accuracy and $x_L=\pi/\sqrt{2}h$ as before.
For $x>x_m$ or $p_{||}<m_E$, screening cannot be ignored to leading
logarithm accuracy. Continuity at $x_m$ fixes
$x_R$, so that
\begin{equation}
F_R=F_L \,\,\frac{\cos\left\{\sqrt{2}\,h\,(x_m-x_L)\right\}}
{\cos\left\{h\,(x_m-x_R)/\sqrt{3}\right\}}
=e^{-\pi/\sqrt{2}h}
\frac{\cos\left\{\sqrt{2}\,h\ln\left(\frac{2\mu}{m_E}\right)-\pi\right\}}
         {\cos\left\{ h\left[\ln\left(\frac{2\mu}{m_E}\right)
           -\ln\left(\frac{2\mu}{F_R}\right)\right]/\sqrt{3}\right\}}
\;.
\label{EX1}
\end{equation}
Eq.~(\ref{EX1}) defines
a transcendental equation for $F_R/2\mu$ as a function
of $N_f,N_c$ and $h$ (equivalently $\mu$), i.e.
\begin{equation}
\frac{F_R}{2\mu}\approx
-{e^{-\pi/\sqrt{2}h}}\,\,
\frac{\cos\left\{\frac{h}{\sqrt{2}}\ln\left(\frac{N_c}{N_fh^2}\right)\right\} }
{\cos\left\{\frac{h}{\sqrt{3}}\ln\left(\sqrt{\frac{N_c}{N_fh^2}}
\frac{F_R}{2\mu}\right)\right\}}\; \; ,
\label{EX3}
\end{equation}
where we have used $m_E/2\mu\approx h\,\sqrt{N_f/N_c}\,\,$, with
\begin{equation}
\frac 1{h^2}\approx \frac{8\pi^2}{N_c g^2}\approx \frac {11}{3}\,{\rm ln}\,
\left(\frac{\mu}{\Lambda_{QCD}}\right)
+\frac {17}{11}\,{\rm ln}\,{\rm ln}\,\left(\frac{\mu}{\Lambda_{QCD}}\right)
\end{equation}
to two loops.
For fixed $N_c$ and in weak coupling ($h\rightarrow 0$), there is no solution
to (\ref{EX3}) as can be seen by inspection. This corresponds to a screening
mass with power suppression, e.g. $m_E/\mu\approx h$.
However, a solution can be found in weak
coupling but large $N_c$, when approximatly
\begin{equation}
e^{-\pi/\sqrt{2}h}\,\sqrt{\frac{N_c}{N_fh^2}}\approx 1
\end{equation}
for which $F_R\approx F_L$. Through $N_c$, this corresponds to a screening
mass with exponential suppression, e.g. $m_E/\mu\approx e^{-\pi/\sqrt{2}h}$.

To assess the minimal value of $N_c$ for
which there is a solution to (\ref{EX3}), it is useful to note that
the solution (\ref{SOLUTION}) is invariant under the shift
\begin{equation}
x\rightarrow x+\ln \left(\frac{\Lambda_{\perp}}{2\mu}\right)
\label{SYMM}
\end{equation}
with similar shifts in the scales $x_{m,L,R}$, implying the existence
of a family of solutions that depend parametrically on
$x_{m,L,R}$ and $\Lambda_{\perp}$. The harmonic equation satisfied by $F(x)$
is scale invariant, hence of the renormalization group type~\footnote{
Indeed, $f(x)=-F'(x)/F(x)$ satisfies $f'(x)=f^2(x)+2h^2$
for $x<x_m$ and $f'(x)=f^2 (x) + h^2/3$ for $x>x_m$, which are the
renormalization group equations derived in~\cite{SON2}, after the
identification $h\rightarrow h/\sqrt{2}$. A similar observation extends to the
BCS case, where $g(x)=-G'(x)/G(x)$ satisfies $g'(x)=g^2(x)+2h_*^2$ (unscreened)
and $g'(x)=g^2(x)+h_*^2/3$ (screened), in agreement with the
renormalization group equations derived in~\cite{SON1}.}.
The scale  $x_R$ is fixed in terms
of $x_{L,m}$ by demanding that the logarithmic derivatives of
(\ref{SOLUTION}) (with pertinent shifts) match at $x_m$. Thus
\begin{equation}
 \frac 1{\sqrt{6}}\tan\left(\frac h{\sqrt{3}}\,
 (x_m-x_R)\right)=\tan\left( \sqrt{2}h\, (x_m-x_L)\right) \; ,
\label{TRANS}
\end{equation}
with
\begin{equation}
x_R=-\frac {\sqrt{3}}h\,\left\{
\arctan\left[ \sqrt{6}\, \tan\left(\sqrt{2} h\, (x_m -x_L)\right)\right] +
{\rm mod}\,\pi\right\}
+ x_m \; .
\label{ROOT}
\end{equation}
The lower bound on $N_c$ or equivalently the upper bound on the
electric mass follows from
\begin{equation}
m_E\equiv \Lambda_{\perp}\,e^{-x_m}=\left(\frac{\Lambda_{\perp}^2}{2\mu}\right)
\left(\frac{2\mu}{\Lambda_{\perp}}\right)\,e^{-x_m}\leq
2\mu\left(\frac{\Lambda_{||}}{\Lambda_{\perp}}\right)\,e^{-x_m}\equiv
2\mu\,e^{-x_{||}-x_m}\,\,,
\label{INEQ}
\end{equation}
where $\Lambda_{\perp}$ and $\Lambda_{||}$ are now exponentially small
scales characterizing the spread in $p_{\perp}$ and $p_{||}$. The inequality
in (\ref{INEQ}) follows from the geometrical constraint $\Lambda_{||}\geq
\Lambda_{\perp}^2/2\mu$ discussed above (see (\ref{PERP}) and also
Fig.~1b).
Up to the rescaling
(\ref{SYMM}), the maximum $\Lambda_{||}$ for which there is a
solution (\ref{SOLUTION}) with positive semi-definite gap,
corresponds to $F(x_{||})=0$,
i.e. $x_{||}=x_R+\sqrt{3}\pi/2h$. 
(The alternative solution $x_{||}=x_L+\sqrt{2}\pi/4h$ does not generate
a maximum bound.) After inserting the latter and (\ref{ROOT})  into
$\hat{c}\equiv\sqrt{2}\, h\,{\rm min}\,(x_{||}+x_m)$, we determine
the minimum as $\hat{c}\approx
2.5051$
and the
lower bound for $N_c$ (upper bound for the electric mass $m_E$)
as
\begin{equation}
\frac{N_c}{N_f} \geq h^2\,e^{\sqrt{2}\hat{c}/h}\; .
\end{equation}
This result is
in overall agreement with a recent renormalization
group estimate~\cite{SON2}~\footnote{After the identification
$h\rightarrow h/\sqrt{2}$ and  $\mu\rightarrow 2\mu$ in the prefactor
of $m_E$
in~\cite{SON2}.}.
In particular, for
$\mu=3\Lambda_{QCD}$, we find $N_c\geq 334\,N_f$.

The case of nonperturbative screening can be addressed similarly
by noting that (\ref{OVERyes}) is now
\begin{equation}
F(p_{||})\approx
\frac {h^2}2\int_0^{\infty}\, dq_{||}\,
\frac{F(q_{||})}{\epsilon_q}\,
{\rm ln}\left(
  \frac{(2\mu\epsilon_q)^2}{(p_{||}-q_{||})^4+m_*^4}\right)\; .
\label{OVERyesx}
\end{equation}
For $x<x_m$ or $p_{||}> m_*$ the screening in (\ref{OVERyesx}) is
inactive. Hence $F(x)=-F_L{\rm cos}(\sqrt{2} hx)$, while for $x>x_m$
or $p_{||}<m_*$ the screening overwhelms the leading logarithm
accuracy with $F(x)={\rm const}$. Continuity at $x_m$ requires
that $x_m=x_L$. Hence
\begin{equation}
m_*=2\mu\, e^{-\frac{\pi}{\sqrt{2} h}}
\end{equation}
which is the maximum tolerated nonperturbative screening mass
for an Overhauser pairing to take place.

Finally, we can qualitatively analyze the effects of temperature
on the Overhauser effect by considering the distribution of
quasiparticles at the Fermi surface. At finite
temperature $T$, the pairing energy becomes
\begin{eqnarray}
F(p_{||})&\approx & \frac {h^2}6\int_0^{\infty} dq_{||}\,
\frac{F(q_{||})}{\epsilon_q}\,
{\rm ln}\left( \frac{2\mu\epsilon_q}{(p_{||}-q_{||})^2}\right)
\,{\rm tanh}\left(\frac{\epsilon_q}{2T}\right)\nonumber\\
&&\mbox{} +\frac {5h^2}6\int_0^{\infty} dq_{||}\,
\frac{F(q_{||})}{\epsilon_q}\,
{\rm ln}\left( \frac{2\mu\epsilon_q}{(p_{||}-q_{||})^2+m_E(T)^2}\right)
\,{\rm tanh}\left(\frac{\epsilon_q}{2T}\right)\,,
\label{ONE3}
\end{eqnarray}
with the temperature dependent screening mass~\cite{LeBELLAC}
\begin{equation}
m^2_E(T)=m_E^2+\left(N_c+\frac {N_f}2\right) \,
\frac{g^2T^2}3\,.
\label{ONE4}
\end{equation}
Even at large $N_c$ the screening mass is finite. We conclude that
at finite temperature, the Overhauser pairing is rapidly depleted
by screening for any value of $N_c$.

\vskip 1.5cm
\centerline{\bf 6. Pairing in Lower Dimensions}
\vskip 1cm

The results we have derived depend on the number of dimensions.
Indeed, the QCD analysis we have carried out when applied to 1+1
dimensions yield the following energy gaps
\begin{eqnarray}
 F(p)& \approx& h^2\int_0^\infty dq\,\frac{F(q)}{\epsilon_q}\,
\,\,\frac 1{(p-q)^2+m_E^2}\; ,\nonumber\\
 G(p)&\approx& h_*^2\int_0^\infty dq\,\frac{G(p)}{\epsilon_q}\,
\,\,\frac 1{(p-q)^2+m_E^2}\; ,
\label{ONE1}
\end{eqnarray}
with the replacement $g^2/8\pi^2\rightarrow g^2/2\pi$ in $h^2$ and $h^2_*$.
Remember that $F(q)$ and $G(q)$ have been defined as even functions.
In deriving (\ref{ONE1}) we have followed the same logic as in 3+1
dimensions, thereby ignoring self-energy insertion on the quark line,
and the gauge-fixing dependence on the gluon propagator. While these
two effects cancel in color singlet states (Overhauser)~\cite{'tHOOFT}, they usually
do not in color-non-singlet states (BCS) except for the case of $N_c=2$~\cite{EBERT}.
In 1+1 dimensions $g^2/2\pi$ has mass dimension, and there is only
electric screening with $m_E^2\approx N_f g^2\,{\rm ln}\,(\mu/g)$.
Clearly,
\begin{equation}
F_0\approx \Lambda \,e^{-{m_E^2}/{h^2}}\gg
G_0\approx\Lambda\,e^{-{m_E^2}/{h_*^2}}\,\,.
\label{ONE2}
\end{equation}
The dominance of the Overhauser effect over the BCS effect whatever
$N_c$, stems from the fact that the Fermi surface reduces to 2 points 
($\pm \mu$) in 1+1 dimensions, with no phase space reduction for the former.
Since both the Overhauser and BCS phase break spontaneously 
chiral symmetry at finite density, the existence of the Overhauser phase 
may rely ultimatly on large $N_c$. The Overhauser effect is dominant in the
Schwinger model where $G_0=0$ because
of the repulsive character of the Coulomb interaction~\footnote{In the
Schwinger model $m_E^2=g^2/2\pi$ independently of $\mu$.},
confirming the results in~\cite{SUSSKIND, SCHAPOSNIK,KAO}.
The case of QCD in 2+1 dimensions will be discussed elsewhere.

\vskip 1.5cm
\centerline{\bf 7. Conclusions}
\vskip 1cm

We have constructed a Wilsonian effective action for various
scalar-isoscalar excitations around the Fermi surface. Our
analysis in the decoupled mode shows that in weak-coupling, the
Overhauser effect can overtake the BCS effect only at large
$N_c$ in the scalar-isoscalar channel, in agreement
with a recent renormalization group result~\cite{SON2}.
The BCS pairing is more robust to screening than the Overhauser
pairing in weak coupling. The BCS analysis was carried out for both
the CFL and the antisymmetric arrangements for arbitrary $N_c\geq 3$,
$N_f\geq 2$, ignoring the superconducting penetration lengths since
the electric and magnetic screening lengths are smaller than the
London and Pippard lengths (for type-I superconductors). In strong
coupling, the Overhauser effect appears to be comparable to the BCS
effect, especially if multiple standing waves are used, allowing
for further cooperative pairing between adjacent patches.
This is particularly relevant for pairings with large energy gaps which
are expected to take place at a few times nuclear matter density~\cite{SchaferWilczek}.

Our effective action is better suited to the use
of variational approximations as discussed in~\cite{DGR92}, and
leads naturally to exact integral equations by variations,
especially in the presence of interactions with retardation
and screening.
It would be interesting to repeat our analysis at nonasymptotic densities
using instanton-generated vertices to address the Overhauser effect. Indeed, for
instantons the cutoff is fixed from the onset by their inverse size.
As we have shown here, the Overhauser pairing, much like the
BCS pairing by magnetic
forces~\cite{SON1}, relies on scattering between pairs in the forward
direction that is kinematically suppressed in the transverse directions
(in fact exponentially suppressed~\cite{DGR92}).
Since the instanton interaction is nearly uniform over the Fermi sphere,
we expect a geometrical enhancement in the BCS pairing in comparison to the
Overhauser pairing. We recall that in the latter the interaction is
enhanced by a factor of order $N_c$. Which one dominates at a few times
nuclear matter density and $N_c=3$ is not clear a priori.
Instantons in the vacuum crystallize for $N_c\geq 20$~\cite{DIAKONOV}
in the quenched approximation, and $3< N_c < 20$ in
the unquenched case. The crystallization is likely
to be favored by finite $\mu$ as the quarks are forced to line-up
along the forward $x_4$-direction.

It is amusing to note that the crystal phase breaks color, flavor,
and translational symmetry spontaneously, with the occurrence of color
and flavor density waves. In many ways, this situation resembles the
one encountered with dense Skyrmions~\cite{SKYRMION} (strong coupling),
suggesting
the possibility of a smooth transition. In the process, color and flavor,
respectively, may get misaligned~\cite{KAPLAN}, resulting into
color-flavor-locked charge density waves in a normal (large gaps)
phase. The Skyrmion crystal at low
density may smoothly transmute to a qualiton crystal at intermediate
densities, with crystalline structure commensurate with the number of
patches on the Fermi surface.
We note that the crystalline structure in $3+1$ dimensions may only show
up as rapid variations in the response functions at momentum $2\mu$.
This is not the case in 1+1 and 2+1 dimensions~\cite{OVERHAUSER}.

Although we have carried out the analysis using Feynman gauge
with minimal changes for the electric and magnetic screening,
we expect our estimates of the gap energies to be reliable
since a close inspection of the equations we derived when reinterpreted
in Minkowski space, shows that the quoted results originate from the forward
scattering amplitude of quarks around the Fermi surface. The latter
is infrared sensitive in the unscreened case and gauge independent,
the exception being in 1+1 dimension~\cite{'tHOOFT,EBERT}.
The similarity between forward
particle-particle and particle-hole scattering
resembles the similarity between forward Compton
and Bhabha scattering. This is what makes $2\mu$ and opposite sides
to the Fermi surface so special between a particle and a hole.

Finally, it is amusing to note that following either the Overhauser or
BCS pairing, the quark eigenvalues of the QCD Dirac operator would suggest
a novel rearrangement that is characterized by novel spectral sum rules.
They will be reported elsewhere. Our use of the effective action at the Fermi
surface is more than a convenience for the study of QCD at large quark chemical
potential. Indeed, given the shortcomings faced by important samplings in
lattice Monte Carlo simulations at finite quark chemical potential, and also
given the importance of Pauli blocking for the non-surface modes, we believe
that a convenient formulation of QCD on the lattice should make use of
Fermionic fields projected onto the Fermi surface, much like the ones
used in the present work, and in the spirit of the heavy-quark
formalism~\cite{PREPARATION}.

\section*{Acknowledgements}

AW and IZ would like to thank Gerry Brown and Edward Shuryak for discussions.
IZ is grateful to Larry McLerran, Robert Pisarski, Dam Son, and Frank Wilczek
for useful discussions. We have benefitted from conservations with Deog Ki
Hong, Steve Hsu and Maciek Nowak. We thank Chang Hwan Lee for help with the
Figure. BYP, MR and IZ acknowledge the hospitality
of KIAS where part of this work was done, and AW the hospitality
of the NTG group at Stony-Brook. This work was supported in part by
US-DOE DE-FG-88ER40388 and DE-FG02-86ER40251.

\end{document}